# Consciousness defined: requirements for biological and artificial general intelligence

Craig I. McKenzie, PhD

Department of Immunology, School of Translational Medicine, Monash University, Melbourne

Correspondence: craig.mckenzie@monash.edu

**Abstract**

Consciousness is notoriously hard to define with objective terms. An objective definition of consciousness is critically needed so that we might accurately understand how consciousness and resultant choice behaviour may arise in biological or artificial systems. Many theories have integrated neurobiological and psychological research to explain how consciousness might arise, but few, if any, outline what is fundamentally required to generate consciousness. To identify such requirements, I examine current theories of consciousness and corresponding scientific research to generate a new definition of consciousness from first principles. Critically, consciousness is the apparatus that provides the ability to make decisions, but it is not defined by the decision itself. As such, a definition of consciousness does not require choice behaviour or an explicit awareness of temporality despite both being well-characterised outcomes of conscious thought. Rather, requirements for consciousness include: at least some capability for perception, a memory for the storage of such perceptual information which in turn provides a framework for an imagination with which a sense of self can be capable of making decisions based on possible and desired futures. Thought experiments and observable neurological phenomena demonstrate that these components are fundamentally required of consciousness, whereby the loss of any one component removes the capability for conscious thought. Identifying these requirements provides a new definition for consciousness by which we can objectively determine consciousness in any conceivable agent, such as non-human animals and artificially intelligent systems.

**Introduction**

The study of consciousness requires the integration of many fields of research including but not limited to neuroscience, psychology, philosophy, physics and artificial general intelligence (AGI).[1,2] Consciousness can be qualitatively assessed by neural pathways,[2-4] but such measurements are limited to biological systems and cannot be defined in a manner that applies to AGI. Definitions of consciousness remain disconnected from the fundamental principles required to generate it.[5] For example, common mistakes include conflating "awareness" with consciousness, likely due to the way the phrase "to be conscious of something" is synonymous with an awareness of that "something". As Crick and Koch wrote in their paper titled "Towards a neurobiological theory of consciousness," they deliberately avoid defining consciousness by explaining that "it is better to avoid a precise definition of consciousness because of the dangers of premature definition."[4] After more than three decades, it is past time we generate a precise definition of consciousness and its requirements that are free of subjective biases.

Defining consciousness accurately is essential for a myriad of scientific and philosophical disciplines. Biologically, it will establish the degree to which other species may experience reality in ways similar to humans. Accurately identifying consciousness is also immediately relevant in cases of disability or trauma that make traditional methods of determining consciousness difficult to interpret (such as testing memory or recognition).[6] Importantly, properly defining consciousness will identify which neuroscientific discoveries are relevant to generating it. In such instances, ignoring what is unnecessary to consciousness may prove to be as helpful in discovering how the brain establishes consciousness than finding what is necessary. At the most ethical extreme, defining consciousness will provide an understanding of whether or not intentional killing of conscious creatures (animal or AGI) is justified. In terms of AGI, defining consciousness allows us to identify the objective goals required to create consciousness through the combination of hardware and software systems and provides a means of determining if an AGI creature is truly conscious. This is paramount to confirm or alleviate fears that conscious AGI will inevitably develop or require suffering in order to be consciouss.[7] Current limitations in our capacity to test for consciousness also necessitate the generation of a precise definition so that consciousness can be swiftly identified should it arise from AGI.[8] At the very least, the ease with which the Turing Test can be passed by unconscious programs demands a need for a clearer definition of consciousness.[9]

In conjunction with the *hard problem* of consciousness proposed by David Chalmers,[10] whereby it remains difficult to explain how biology can give rise to subjective experience, fundamental principles of consciousness can be considered as a *hard line* of consciousness in order to establish whether or not a creature (biological or artificial) is genuinely conscious. Those on one side of the hard line are unconscious, those on the other are conscious. As such, the hard line is to be a precise definition of consciousness and thereby outline the requirements for its existence or creation. In the current paper I draw the hard line by arriving at a definition of consciousness based on first principles. I use experimental and theoretical evidence to determine the fundamental principles of consciousness which are required to 1) confirm the presence of or 2) create consciousness. I discuss the literature surrounding current theories of consciousness and provide an analysis from which a precise definition of consciousness must arise.

**Theories of consciousness, their strengths and weaknesses**

Current definitions of consciousness fall under a wide umbrella of categories, many of which overlap (**Table 1**). Broadly, the most interrogated theories of what consciousness is or how it arises are integration, global workplace, first-order and higher-order theories.[11,12] In **Table 1**, I summarise current and past theories of consciousness to highlight their strengths and faults in order to consider the fundamental requirements for consciousness. These theories and their problematic terminologies have also been reviewed recently by Riccardo Fesce and the difficulty of their unification into a theory of consciousness is aptly discussed.[5] Such theories are more usefully interpreted as potential mechanisms by which conscious systems can operate, rather than strict definitions of consciousness per se. This has been demonstrated in a recent whitepaper exploring the correlates of consciousness in artificially intelligent systems that outlined 14 different theories of consciousness and used these theories as a means of assessing consciousness in current AI large language models.[13] However, in the absence of a precise definition of consciousness based on fundamental requirements we remain limited to determining correlates of consciousness, rather than truly identifying consciousness itself.

**Table 1. Theories of consciousness**

| Theory | Strengths | Weaknesses | Lessons for defining consciousness |
|---|---|---|---|
| **Integration** Information Theory (ITT) defines consciousness based on differentiation and integration of thought that leads to effective discrimination from a large number of possible states. It postulates that | This is likely a valuable way of understanding the biology of how cells in the brain can generate consciousness. The axioms of IIT of consciousness provide a basis for understanding fundamental (axiomatic) | Axioms outlined so far in IIT do not address all facets of experimental or theoretical consciousness widely identified by research.[16] Furthermore, like discrimination theories, the axioms of IIT suffer from defining | The very existence of consciousness suggests there must be axiomatic principles that determine it, and although the axioms of IIT have not sufficiently described consciousness, they highlight that we must focus on identifying such |
| consciousness arises from a series of logic gates in the physical substrate of the brain.[14] | principles that may define consciousness[14,15] | consciousness by its output, rather than defining the fundamental components required to generate that output. | first principles of consciousness if we wish to understand it. |
| **Global workplace** theory (GWT; or **access** theory) refers to how subjective experience, considered a core component of consciousness, arises from a myriad of neuronal networks that work together or compete to ultimately give rise to the active neurons responsible for such experience.[17-19] Consciousness arises from accessing only relevant perceptive information from a larger pool of perceptive information. | GWT is valuable when considering how consciousness might arise from neural networks by highlighting that what we are unconscious of is crucial to establishing consciousness.[20] | Similar to IIT, this theory avoids detailing how or which neuronal pathways lead to subjective experience. | There is more to defining consciousness than subjective experience. Rather, a theory of consciousness must include how such experience arose. To address this, I contend to be consistent with GWT by defining the fundamental requirements for consciousness as that which lead to subjective experience and particularly choice behaviour. Furthermore, GWT also infers that perception is critical for consciousness, a valid principle upon which to define it. |
| **First-order** theories posit that conscious states arise from awareness (or sensory representations) without requiring additional processes such as attention or working memory to feed back into said representation.[21,22] | Describing consciousness by accessing representations is an intuitive model for understanding consciousness. For example, upon seeing a face, a representation of the face is what the conscious mind experiences. | It remains difficult to scientifically prove that only first-order awareness is necessary for consciousness whilst so many neural pathways appear to be involved. It has been observed that some first-order representations may arise outside of consciousness.[23] | This theory focuses on the subject experience of consciousness through sensory representations. It highlights how defining consciousness should at the very least address how subjective experience can arise. This is a central building block upon which other theories of consciousness can expand. |
| **Higher-order** theories (HOTs) define consciousness by describing how consciousness can exist outside the scope of awareness. Such theories suggest that one can have conscious processes that are not included in subjective awareness whereby there are higher-order conscious thoughts that are not necessarily registered by the creature experiencing them.[24] | Like other theories, we can glean from HOTs that at least some capacity for perception is required for consciousness by generating the required awareness. In the case of blindsight, HOTs contend that conscious thoughts can recognise objects without the blinded subject being aware of them.[23] | Claims that consciousness extends beyond subjective experience are difficult to validate. Thoughts that fall beyond subjective experience are likely subconscious processes – like the parasympathetic nervous system – rather than higher-order states of consciousness. Although actions in individuals with blindsight are remarkable, so too are other subconscious actions like breathing and blinking in response to the changes in our environment. | Rather than consciousness existing outside the scope of awareness, it is more likely to rest entirely within it, as per GWT. This is because consciousness likely gives rise to subjective experience and decision making, and nothing else. All other processes are subconscious, which by convention should not be used to determine a definition for consciousness. |

The content of consciousness has been described as 1) subjective phenomenal experience, 2) a seemingly transparent experience of reality and 3) experience of consciousness in the present, yielding a sense of presence.[25] Further divisions include the distinction between subjective phenomenal consciousness and that of access consciousness in which representation of reality can be confirmed by verbal content.[26] Though these accurately describe the output of consciousness like many theories mentioned above, they do not describe what is required to generate consciousness. Such a problem is highlighted when considering the neural correlates of consciousness (NCC) whereby it is difficult to determine whether a neuronal pathway is generating consciousness or is an outcome of consciousness.[27] It can be unclear if correlates are a cause or effect of consciousness.

To explore consciousness further by thought experimentation, Thomas Nagel has asked what is it like to be a bat?[28] This describes consciousness as having a relatable property between differing minds, in which we can imagine what another conscious mind can 'be like'. For example, we can imagine what it is like to be a bat so it is likely conscious, we cannot imagine what it is like to be a window so it is not conscious. This method of thinking about consciousness raises difficulties for defining consciousness because it is describing something by what it is not and is underpinned by subjective experience. It is attempting to describe consciousness by comparing it to what is not conscious. Any definition of consciousness must therefore avoid comparisons between creatures or objects and instead should provide a single definition into which both could be categorized (conscious vs. not conscious) without a need to compare them in the definition itself.

Further problems in defining consciousness have arisen from relying too heavily on human neurobiology or psychology. For example, consciousness has been described as having "an extraordinary range of different contents—perception in the various senses, endogenous imagery, emotional feelings, inner speech, concepts, action-related ideas and 'fringe' experiences such as feelings of familiarity".[29] These properties of consciousness would suggest that a major component of defining consciousness can come from comparing it with human experience or biology. Although this is crucial in understanding how consciousness works in humans and other animals, it may be a misleading path to follow if one were to try and define consciousness by such comparisons. As is outlined below, emotions might influence human decisions but they are unnecessary to the capability of making a decision – and are thereby unnecessary for defining consciousness. Conceivably, it will be possible in the future to construct consciousness without the need for biological components, and as such it is important that we can define consciousness without comparisons, but rather hard objective terms.

*The problem with emotion*

Emotion and feeling are often described as a quality of consciousness but many emotions have been described as subconscious.[30,31] The most widely researched example of subconscious emotion is fear, whereby autonomic responses consistent with fear need not arise from being consciously aware of the fear stimuli.[32-37] Neural activity in the amygdala suggests that initial behaviour from sudden fear is a subconscious response, whereas attentional responses arising sometime after the initial fear stimulus are a conscious process.[38] As such, consciousness appears to be reactionary to the emotion of fear. We can therefore posit that such feelings or emotions are not part of the apparatus that creates our consciousness. Feelings, like senses, are something we experience in the present and are what our consciousness uses to make decisions. Emotions are our perception of our internal landscape whereas our senses perceive the external one. As such, our consciousness is the apparatus that decides what we do with our senses or feelings rather than our senses and feelings arising from conscious decision. Feelings can alter our conscious thoughts and our thoughts can alter our feelings, but one is not required for the existence of the other. For example, many outcomes of feeling fear are subconscious processes, such as a quickened heart rate, jumping in fright or sudden weakness. Stimulation of fear networks in animal brains can generate what appear to be feelings of fear, but this does not enhance or detract from the animal's potential for conscious thought due to the consistent behaviours exhibited in response to such fear.[39]

*Temporality of consciousness*

Some definitions for consciousness stretching back over a century to William James involve a requirement for the sequential temporal order of consciousness and a continuity of the perception of time.[40-42] In effect, continual slices of present information generated from the senses become memorised and create a sense of the past in context to the ever-updating present. This temporality should not be included as a fundamental tenet of consciousness because our sense of temporality arises from our perceived reality rather than arising from consciousness. The fact that time moves in one direction, albeit at different relative speeds, ensures that a conscious executive will order precepts temporally. It is physically impossible for a conscious creature to experience the future and remember it as past because time does not ever skip backwards, however, if it did our consciousness would arrange our memories as such. Sense of temporality does not define consciousness but, like choice behaviour, is an outcome of it. One should note that memory or imagination might approximate what has or will happen, but cannot actually generate an experience of the actual past or future that has occurred or will occur.

Our perception of time is inaccurate, thereby making it difficult to incorporate temporal perception into a definition of consciousness. Perceptive information reaches our brain milliseconds before the conscious executive can experience them.[43] As such, our perceived present is delayed from reality. Conscious awareness of one's own decisions is also delayed milliseconds after such decisions become inevitable.[44] This delay between reality and perceived reality ensures that incorporating the passage of time into models of consciousness is problematic. However, as perceived reality is the conscious reality, we can avoid this difficulty by ensuring that any reference to temporality is relative to the conscious mind experiencing it. As such, the present time can be considered to be the moment that the conscious mind becomes aware of percepts or decisions, rather than the explicit moment that the experience of such percepts or decisions is occurring.

**Building blocks of consciousness**

*Perception and memory*

Sensory information that informs our perception of the present, such as our five major senses, is critical in understanding how consciousness might be defined but is not to be considered indistinguishable from consciousness itself. Rather, sensory information informs our consciousness on what is an approximation of reality. For example, the idea that consumption of a psychedelic substance that provides the perception of seeing more vivid colours makes one "more conscious" is untrue, as "higher" states of consciousness are not achieved by such psychedelic drugs.[45] Indeed, associating perception with a sliding scale (higher or lower) consciousness is misleading. It is akin to suggesting a visually-impaired person is less conscious than a person of normal vision, which is certainly not the case. As such, the accuracy of one's approximation of reality through sensory input does not determine whether or not one is conscious. For example, if time flowed differently, no doubt our consciousness would have evolved without a need for experiencing a sequential passage of time. As such, the fundamental laws of the universe and the flow of time determine the constraints of what consciousness perceives, rather than the other way around. Requirements for consciousness discussed in this paper must therefore generate these contents of consciousness within the bounds of perceived reality, but not be defined by what it is being perceived.

One must consider the accuracy of perceptive information with regards to objectively verifiable reality. Getting more accurate sensory information (approximating reality) may allow us to make better decisions, but it does not enhance our consciousness. As such, the extent to which we understand the truth of the universe has no impact on whether or not we are conscious. The accuracy of our senses (perception) informs our consciousness with the true context in which we exist in past, present and possible future but does not provide a basis for consciousness itself. Senses therefore provide the basis of consciousness independent of the accuracy of such senses.

*Subconscious needs give rise to conscious desires*

Evolution has provided a selective pressure for biological systems to excel at successfully meeting their needs within a given environment. This gives rise to the biological adaptations that help us survive and reproduce. In the case of creatures capable of feeling emotions like mammals, this same pressure has given rise to such phenomena as hunger and thirst as a means of meeting our needs. The biological need drives the feeling, whereby a lack of food drives hunger, and a lack of water drives thirst. This provides a means by which needs can drive behaviour, as the hungry seek food and the thirsty seek water.

Needs do not arise from conscious thinking, rather they arise subconsciously. The apparatus that drives these needs and their subsequent behaviour evolved before consciousness. This includes rudimentary neural systems that can mediate the pursuit of the creature's needs such as a bacterium's use of a flagellum to increase the likelihood of infecting a host.[46] It is highly likely that these initial subconscious behaviours provided the biological apparatus for consciousness to evolve. To pursue one's needs, a vast array of biological systems must be coordinated, such as moving toward water and being coordinated and capable enough to drink it. Consciousness can therefore arise once such a system develops because it goes on to further optimise the pursuit of needs by providing desires for choice behaviour. Choice behaviour then provides the opportunity for an organism to act beyond their needs, and instead is driven by personal preference as per the conscious decisions that drive it. As such, the needs of our non-conscious ancestors provided an evolutionary pressure to select for creatures that best achieve those needs, and subsequently provided a stepping stone for the evolution of consciousness.

We can conclude from this that desires are driven but what a conscious biological creature wants, rather than precisely what they biologically need. Consciousness therefore goes beyond the optimisation of meeting one's needs and provides an opportunity to spend one's energy on other things of their choosing. Desire therefore arises as a certain kind of temporary or contextually relevant need in the conscious mind, becoming the primary motivator of choice behaviour. Desire is therefore a critical component of consciousness, independent of the choice behaviour it ultimately mediates.

*Consciousness, dreams and imagination*

How sleep and dreams affect consciousness can also aid in the attempt to objectively define it. Importantly, to experience dreams with a conscious mind is a conscious experience. This was postulated by Crick and Koch in 1990 and evidence for this arises from measuring the brain waves of sleeping macaques, whereby gamma-band activity during random eye movement (REM) sleep was similar to gamma-band activity generated whilst they were awake.[4,47] Studies of sleep reveal that a dreaming mind is conscious despite the inaction of the real-world body.[48-50] In contrast, brain waves from macaques in non-REM sleep or sleep induced by an anaesthetic do not exhibit the same patterns as from REM sleep or wakefulness.[47] This suggests dreamless or anaesthetic-induced sleep does not involve a conscious mind. The mind has crossed back over the hard line of consciousness from conscious to unconscious.

A conscious mind experiencing a dream is not aware of reality but instead becomes aware of a non-real, imaginary world generated by their own brain. Despite being conscious and capable of choice, a dreamer does not generate choice physical behaviour from dreamt decisions when compared to the same decisions when made awake. A dream may cause small but perceptible movement in muscles associated with dreamed activity,[51,52] but sleep has provided sufficient regulation to completely restrict the vast majority of physical behaviour. Therefore, the conscious mind in a dreaming state proves that dynamic choice behaviour is not required for a definition of consciousness. Instead, consciousness is required to generate the choice that could lead to behaviour, but is not contingent upon the behaviour itself. This further suggests that using behaviour or accurate report as a means to measure consciousness is insufficient for determining whether or not a creature is truly conscious. Rather than behaviour, it is the capability to generate decisions in a dreaming state that is more reflective of consciousness. When awake, such decisions can lead to behaviour in the real world. When dreaming, such decisions lead to imagined behaviours in an imagined world but these decisions still arise from conscious thought.

To further integrate lessons from dreaming into a definition of consciousness, I posit that dreams arise as a combination of memory and imagination to build real percepts of non-real scenarios. Such scenarios are usually a mixture of familiar and new, whereby the "familiar" involves memories whilst the "new" comes seemingly at random from imagination or long-term memories that have not been consciously remembered. This may be a process by which our brains practice conscious decision making. Dreams present our consciousness with new scenarios that maintain some kind of basis in reality (from memories) and something unfamiliar (from imagination). As such, our consciousness is allowed to make decisions in a scenario without real consequences – a kind of 'practice' for real world decisions. Underpinning such conscious dreaming is therefore the imagination of potential actions, highlighting how imagination is critical to consciousness and the pursuit of desires – even in a hallucinated reality such as a dream.

When a dreamer does not anticipate what is being dreamt about, the conscious self that experiences such dreams has maintained sufficient control of imagination to act on desire but has lost control of the imagination that creates the hallucinated perceptual information experienced as a dream. In this instance, subconscious imagination is used to generate the external environment whilst conscious imagination is maintained so that decisions remain possible. However, when the dreamer can anticipate or even control such dreams in what is considered 'lucid dreaming',[53] the conscious mind experiencing the dream has remained (at least partly) in control of their imagination for both their own actions and of their external environment of the dream. Being in control of imagination therefore leads to control of the dream, because the dream is being constructed from imagination. This is likely how dreams arise in which the dreamer can control their experience, such as having control over non-real capabilities like flight. This suggests once again that consciousness is independent of the reality of one's perception, for even when such perceived reality is consciously controlled does it fail to alter the capacity for consciousness itself.

Although consciousness can be observed in dreams, I would like to make the distinction that consciousness is not a requirement for dreaming. This is likely the case when foetuses have been observed to reach REM cycle sleep in which their minds appear to be dreaming.[54,55] Without developing a considerable memory (as displayed by a lack of understanding object permanence in infants) it is probable that foetuses are not conscious when they dream. Likely their minds are presented with dreams of being in the foetus, with all the sounds and sensations that may bring, but the mind experiencing it has little to no memory to give it context in past or present. As such, their dreams are likely constant streams of percepts that, though they might trigger emotions from subconscious responses, have no long-term meaning whatsoever to the yet-to-be-conscious mind that experiences it. This suggests that consciousness may arise much later in human development than commonly thought.

**Requirements for consciousness**

The relationship between memory, perception and consciousness has been interrogated since at least the 4th century, in which St. Augustine aptly discussed their relationship in *The Confessions*:

*Perhaps it might be said rightly that there are three times: a time present of things past; a time present of things present; and a time present of things future. For these three do coexist somehow in the soul, for otherwise I could not see them. The time present of things past is memory; the time present of things present is direct experience; the time present of things future is expectation.*[56,57]

Here I will discuss the requirements for consciousness that include and extend beyond perception, memory and imagination (or St. Augustine's 'expectation') to integrate sense of self and desire into a framework that defines consciousness by first principles. Having discussed various evidence for the requirements of consciousness, I can distil here such requirements without the pitfalls of prior theories. Such requirements are as follows: perception, memory, imagination, sense-of-self, desire and decision.

**Perception** involves the use of senses to approximate reality. It must be stated that in many instances our human senses do not reflect reality, but rather an approximation of reality that has been evolutionarily selected to improve biological fitness.[58] However, regardless of the accuracy of perception, at least some perception of the surrounding reality is required for a conscious mind to generate memory, imagination and a sense of self. Perception can generate these fundamental components of consciousness by providing a means with which the conscious mind can discriminate between what is self and non-self. It can do so without requiring the totality of perceptive capabilities available to it, thereby generating an awareness from an overflow of perceptive data.[59] That humans can see more than they remember is strong evidence to suggest that at least some perceptive information is required for consciousness.[59,60] Indeed, the lack of perceptive capability in disembodied neural organoids grown in a laboratory ensures such organoids remain incapable of generating representations of meaning, and thereby incapable of consciousness.[61]

Attention is an output of a conscious process that arises from at least some capacity for perception, providing a further argument that perception is another fundamental requirement for consciousness despite not being part of the conscious process.[62] The cognitive cycle proposed by Baars and Franklin supports this claim, whereby perception is required to generate working memory accessed by the conscious executive.[63]

As intimated by Charles Richet over a century ago, "without memory no conscious sensation, without memory no consciousness".[64,65] **Memory** is defined here as the storage of perceived information. In the case of biological creatures, this information is generated by the senses but in the case of artificial intelligence memory could conceivably be enhanced by access to digital data storage. Such information must at least contain 1) percepts (from senses originating from self) and 2) experiences (from putting percepts into context to one's self). Working memory theories of consciousness have established that memory is required to generate an understanding of what is self and non-self by providing stored information on an approximated internal and external reality. Without a databank for such reality, there can be no output of conscious thought. This is highlighted by the conscious copy theory which suggests that although some working memories can have an impact on behaviour without consciously being remembered, there must be at least some capacity for memory to generate conscious introspection.[65]

NCC have demonstrated a purportedly profound role for **imagination** in generating conscious decisions which highlight imagination as a requirement for consciousness.[5] Imagination is defined here as an ability to think critically and creatively in a manner that mentally represents that which is not presently perceived.[66] Specifically, critical or logical thinking is the ability to comprehend or predict the consequences of actions or inaction from what is already known. This is a non-random process. In contrast, creative thinking is the ability to comprehend or construct non-real possibilities that can be experienced as percepts without generating actual percepts. In effect, creativity from imagination generates non-real memories (imagined futures) from possibilities. Such creativity is a partially random process whereby entirely creative thoughts can arise without a basis in memory. A number of real-world examples can demonstrate these principles of imagination and their relation to conscious thought (**Table 2**).

**Table 2. How imagination generates critical thinking and creativity that underpin conscious thought.**

| |
|---|
| Example 1: The partial randomness of creativity is akin to random mutation in evolution that generates new genes that go on to create new phenotypes. In this example, the randomness component is the mutation in the gene but the non-random component is the selection pressure that determines whether those new genes are passed on or not. For example, too much randomness in creative art ensures that whatever created it largely loses meaning to most people. Too little randomness and it is akin to logic, whereby a problem is solved entirely based on prior knowledge (as opposed to creating new processes to discover new knowledge). |
| Example 2: An example of the difference between logic and creativity is mathematics. Following an equation to solve a problem is logic. To generate a new equation to solve a previously unsolvable problem is creativity. Both are imaginative processes that demonstrate consciousness. |
| Example 3: Remember what you did on your last birthday. Now imagine that a 100ft-tall elephant wished you "happy birthday". Neither your memory of the birthday or the imaginary elephant are real in the present time, but they are comprehended or envisioned by your conscious mind in a similar manner. Both rely on your memories of your birthday and elephants. This demonstrates how conscious imagination and memory are intertwined with perception, regardless of the reality of such percepts. |

A **sense of self** is commonly considered a requirement of consciousness.[67-69] It provides a conscious creature with the capability to discriminate between self and non-self, which subsequently provides an understanding of how the self can manipulate the non-self to generate behaviour from decisions. It provides the capability of a creature to understand 1) what it is and is not and 2) what it does and does not desire. A sense of self should not be conflated with self-importance or the auto-biographical self. Rather, this sense of self is determined by successfully discriminating between self and non-self in a manner that may be described as 'intrinsic' and thereby inferred by – but not necessarily dependent upon – an awareness of one's physical body.[70] The self is that which the conscious mind can control and that no other conscious entity can likewise manipulate directly. The non-self is that which a conscious mind cannot control with conscious thought alone. For example, someone that picks up an inanimate carbon rod is physically in control of it, but the rod is not part of their consciousness because it cannot be manipulated in the absence of physical interaction. Another example would be that a hand is controlled by the self but a hair is not, because hair cannot be controlled with conscious thought whereas the hand can be consciously controlled to cut the hair.

It has been suggested that there need not be a sense of self within a definition of consciousness, and that evidence for this is seen when one meditates or hallucinates to a point of losing all awareness and yet retaining a primordial sense of attention.[70] However, this sense of selfless attention is unlikely to be a state of consciousness because losing the sense of self renders attention pointless to the mind experiencing it. Decisions, and therefore actions and choice behaviour, cannot be made without a sense of self.

The ultimate output of waking consciousness can be considered to be choice behaviour whereby conscious thought generates physical actions. This is distinct from forced behaviours such as those that are "hard-wired" by genetic code or the programming in software that drives the hardware of a robotic arm. An example of such pre-programmed behaviour can be found in squirrels raised in captivity from birth in that they still attempt to bury nuts despite never seeing such behaviour and living in a cage unsuitable for digging.[71] As discussed above, what must be stressed is that consciousness is capable of generating behaviour – but does not require behaviour to be defined as conscious. Behaviour is not a requirement for consciousness. Rather, the capability for dynamic behaviour driven by imagination is an *outcome* of consciousness. As previously mentioned, the lack of behaviour despite consciousness is observed most strikingly in dreamers, whose minds remain conscious of an imaginary world despite their bodies being largely inactive. Such choice behaviours are downstream of **decision** making so it is therefore more accurate to consider decisions (or the 'choice' itself) as a fundamental requirement of consciousness. All choice behaviours are generated from decisions, but not all decisions generate choice behaviours.

But from where do these conscious decisions stem? In a straightforward cause-and-effect, decisions arise from **desire**. Due to the requirement for decision-making to be a fundamental component of consciousness, it suggests that desire is also a fundamental component because without desires one cannot have any impetus to make decisions. Desires can arise from a hardwired source, such as thirst, lust and hunger, which are generated from innate needs. Desire is therefore the point at which sub- or non-conscious processes within the brain can generate conscious thought. Although desires arise independently of consciousness, their manifestation is a requirement for consciousness.

**Defining consciousness**

The arguments presented above can be used construct a definition for consciousness from its fundamental requirements. Here I outline how these requirements relate to one another in sequence.

Consciousness requires 1) perceptive capabilities (such as the senses) to provide an approximation of the present reality in order to generate 2) memories of past experience, which in turn provides an understanding of 3a) how the reality of the present came to be and 3b) the imagination of possible futures. The capability to distinguish 4) self and non-self amongst these memories and imaginings provides a sense of self through time. In turn, this leads to a capability to 5) make decisions in the present based on 6) internal desires that provide comprehension of real or imaginary past and future realities (3a) and ensure imagined futures (3b) become reality through choice behaviour. As a theory of consciousness must incorporate the output of cognitive function,[72] it is important to state that consciousness must generate decisions, which in turn are responsible for choice behaviour in real or imagined (dream) realities. These requirements and outputs of consciousness are connected in a network that is distinctly separable into the human experience of past, present and future (**Figure 1**).

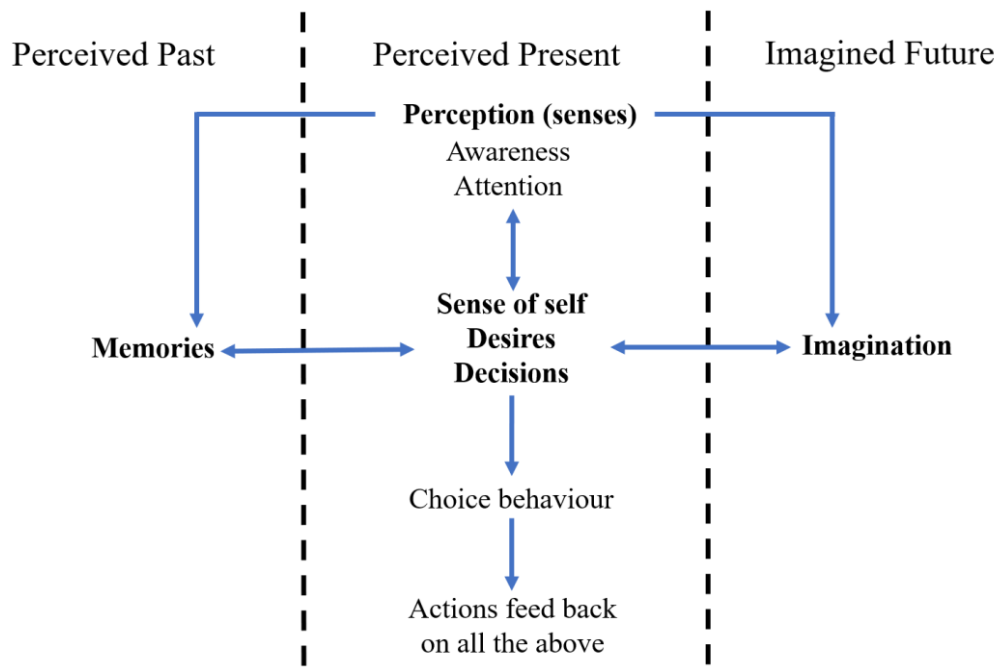

**Figure 1. Model network of consciousness integrated into the experience of time.** Perception generates memory of an approximately 'real' past and provides a reasonable framework for an imagination of potentially realisable futures. The combination of memory and imagination with a sense of self, creates the opportunity to generate decisions that satisfy desires. These decisions lead to choice behaviour. Bolded words are fundamental requirements for consciousness.

Taken together, we can combine our understanding of these requirements for consciousness into a new definition of consciousness:

*Consciousness is the capacity to generate desires and decisions about perceived or imagined realities by distinguishing self from non-self through the use of perception, memory and imagination. Perception generates memory (associated with awareness, attention and experiences) and imagination (critical and creative thinking), leading to the formation of decisions based on desires that are not directly hardwired by preprogrammed code (such as genes or software). These conscious decisions are necessary for choice behaviour and provide the capacity for a conscious agent to adapt to perceived or imagined realities.*

**Thought experiments to interrogate the requirements of consciousness**

A core component of biological experiments is to remove a single variable (for example, abrogating the function of a single gene in mice to determine the gene's function) and observe what that variable was responsible for. To do so for the above definition of consciousness, I will remove one component of the definition at a time in order to theoretically prove their fundamental importance for conscious thought.

*Removing perception*

Without any ability to perceive one cannot distinguish self from other nor be capable of storing any information in memory that might be used to generate decisions. If all perceptive information were absent from a creature from the beginning of its development, it would therefore be impossible to develop an understanding of internal or external environment and have no reference for subjective experience. Perceptive information is therefore a foundational pillar for consciousness. As I have intimated earlier, simply a capacity for at least some perception is required for consciousness, as consciousness can arise regardless of whether the perceptual information is relatively small or inaccurate compared to reality.

*Removing memory*

Without memory, creatures would exist in a constant state of imagining possibilities without understanding any context as to what is possible. This is not consciousness as such creatures lacking any memory will be stuck imagining things over and over again when stimulated by the same percepts. For example, looking at a chair will trigger an imagination about what such an object is and what it could be, but these imaginations never have an end point because without memory they are immediately forgotten and either repeated or replaced. Such a memoryless creature is analogous to the mobile robots described by Rodney Brooks' whereby learning algorithms applied to continuous input of perceptive data allow the robot to move about without a memory.[73] It should be noted that though these robots were described as "intelligent," they are not conscious.

*Removing imagination*

Similarly, without imagination there can be no consciousness even in the presence of memory. With no imagination and only memory, creatures can build a sense of self from the past but can never imagine a sense of self into the future. At this stage, all actions of such creatures would be based on what has been learned. Such actions would be instinctual and could not anticipate situations which they had not already experienced or been taught about. Interestingly, with the exclusion of physical pain, it may be impossible for such creatures to mentally suffer because they cannot imagine having a better situation than they currently possess.

*Removing a sense of self*

Imagine a creature without a sense of self. In effect, this creature is permanently experiencing Ego Death. With a memory and imagination, it can learn or understand. However, without a sense of self there can be no distinguishing between self and non-self and such a creature has no way in which it can generate desires or decisions and therefore cannot generate choice behaviour. Taking conscious actions requires thoughts to direct the self, so without a sense of self there can be no action. This is not to say that all actions are conscious. Many actions are subconscious, such as those controlled by the parasympathetic nervous system (like breathing) or behaviours determined by genetics (like squirrels burying acorns). However, conscious decision making can only lead to choice behaviour if a sense of self is identified and the body instructed on how to act by desires and decisions.

**Conclusion**

The difficulty with defining consciousness has stemmed from difficulty in determining precisely what is required for consciousness to exist. Theories have run aground by defining consciousness by what it is not, by similarities to humans or by failing to address resultant choice behaviours of conscious creatures. Here I have proposed a set of requirements for consciousness that provide a hard definition for its existence: memory, imagination, sense of self and a capability for perception that informs, rather than controls, the desires and resultant decisions generated by a conscious mind. As such, senses and especially feelings are separate from what can be considered consciousness, yet both can influence the decisions generated by such a consciousness.

Currently, with the rise of artificially intelligent language prediction models like OpenAI's Chat GPT, it is conceivable that genuine AGI may develop once such digital systems are provided needs or desires given their already remarkable capacity for perception, sense-of-self and memory coupled with an imagination beyond what was already programmed by their digital code. Given the above definition of consciousness AGI could be achieved by providing current artificially intelligent models with a sense of self and defined desires in conjunction with a sufficient method for storing the digital memories of what it has done such that it could imagine what it could do.

Further applications of this definition of consciousness can be applied to biology, such as determining consciousness in cognitively impaired individuals, non-human intelligence or within cells or tissues of organismic systems. For example, the adaptive immune system can be considered conscious. Such a collection of cells can detect and interact with antigenic stimuli due to its perception, memory and imagination of antigenic shapes (through long-lived cells in combination with recombination and somatic mutation of antigen receptors). Given that the adaptive immune system undergoes a tolerance to 'self' such that it can identify 'non-self', the adaptive immune system is capable of consciously deciding how to respond to new antigenic information. Rudimentary consciousnesses such as those in the immune system can demonstrate that consciousness may arise independently outside of neural networks if selective pressures are sufficient to do so.

Perhaps the best example for these first principles of consciousness originates from pre-history. Remarking on the discovery a 44,000-year-old Indonesian cave painting, Archaeologist Maxine Aubert saw "the ability for humans to make a story, a narrative scene, as one of the last steps of human cognition".[74] It is the imagination of this pre-historic painter that provides us with an understanding that the painter was conscious. The artist's depiction of humans hunting buffalo is much more than something that simply no longer exists. It *did* exist or *could* exist. It demonstrates memory and imagination, both of which were expressed in the choice behaviour of an ancient human with the sense of self required to express it.